\documentclass[aps,prl,twocolumn,showpacs]{revtex4}
\usepackage{graphicx}

\begin{document}
\title{Critical Behavior of Frustrated Josephson Junction Arrays with Bond Disorder}
\author{Young-Je Yun}
\author{In-Cheol Baek}
\author{Mu-Yong Choi}
\affiliation{BK21 Physics Division and Institute of Basic Science, Sungkyunkwan University,
Suwon 440-746, Korea}
\date{14 March 2002; published 28 June 2002}

\begin{abstract}
The scaling behavior of the current-voltage ($IV$) characteristics of a two-dimensional proximity-coupled Josephson junction array (JJA) with quenched bond disorder was investigated for frustrations $f=1/5$, $1/3$, $2/5$, and $1/2$.
For all these frustrations including 1/5 and 2/5 where a strongly first-order phase transition is expected in the absence of disorder, the $IV$ characteristics exhibited a good scaling behavior.
The critical exponent $\nu$ indicates that bond disorder may drive the phase transitions to be continuous but not into the Ising universality class, contrary to what was observed in Monte Carlo simulations.
The dynamic critical exponent $z$ for JJA's was found to be only 0.60 - 0.77.
\end{abstract}


\maketitle

The addition of bond randomness may change the first-order transition into the second-order one.
A particular manifestation of the rounding of the first-order transition is seen \cite{R1,R2} in the two-dimensional (2D) $q$-state Potts model, for which the transition in the absence of disorder is of first-order for $q>4$.
For the eight-state Potts model, the critical exponents of the disorder-driven second-order transition were found \cite{R3} to fall into the universality class of the 2D Ising model.
Kardar {\it et al}. \cite{R4} suggested that the disorder-induced Ising criticality is due to the asymptotically Ising-like critical interface.
It has been shown in the recent Monte Carlo (MC) simulations \cite{R5} that bond randomness has a similar effect on the phase transition of the uniformly frustrated 2D $XY$ model.
When quenched bond disorder was introduced, the first-order transition of the $XY$ model at frustration, the fraction of a flux quantum per plaquette, $f=2/5$ was found to change into the continuous transition with the critical exponents of the 2D Ising values.
Despite the difference in the underlying symmetry of the $XY$ model from the $q$-state Potts model, the disorder-induced Ising-like transition at $f=2/5$ was also attributed to the Ising-like domain wall excitations.
The numerical observations were claimed to be consistent with the critical exponents obtained experimentally \cite{R6}.
However, because of a systematic error in locating $T_c$ of the sample, a superconducting wire network, the announced error bar for the experimentally obtained critical exponents is $\pm 0.5$, somewhat too large for identifying the Ising universality class.
It appears to us that an experimental confirmation of the Ising criticality at $f=2/5$ with disorder has yet to be done.
Another interesting question raised by the numerical observations at $f=2/5$ is whether the transitions at different frustrations such as 1/5, 1/3, and 1/2 can also be driven to the same Ising universality class by bond disorder.
In the absence of disorder, expected numerically are a strongly first-order transition at $f=1/5$ \cite{R7}, an Ising-like transition at $f=1/3$ \cite{R5}, and a continuous but non-Ising-like transition at $f=1/2$ \cite{R8}.

Proximity-coupled Josephson junction arrays (JJA's) and superconducting wire networks (SWN's) exposed to a uniform transverse magnetic field are the closest experimental realizations of the uniformly frustrated 2D $XY$ model.
In JJA's and SWN's, the phase transition appears in the form of a superconducting-to-resistive transition associated with the melting of a vortex lattice.
If the superconducting transition is a continuous one, the critical behavior of the phase transition can be investigated experimentally by examining the scaling behavior of the current-voltage ($IV$) characteristics of the superconducting sample \cite{R6,R9}.
For the purpose of determining the critical exponents, JJA's gain an advantage over  SWN's.
A broad critical region of the JJA allows one to determine the critical exponents with accuracies sufficient for identifying the universality class.
We investigated the scaling behavior of the $IV$ characteristics of a square proximity-coupled JJA with the deviation of the junction coupling strength $\lesssim$ 15 \% at four different frustrations $f=1/5$, $1/3$, $2/5$, and $1/2$.
The scaled $IV$ data reveal that the phase transitions at all these $f$'s are continuous, but none of the transitions including those at $f=2/5$ and 1/3 belongs to the Ising universality class.
It was also found from the scaling analyses that the dynamic critical exponent $z$ of the JJA's is only 0.60 - 0.77, much smaller than that observed in SWN's.

\begin{figure}
\includegraphics[width=1.0\linewidth]{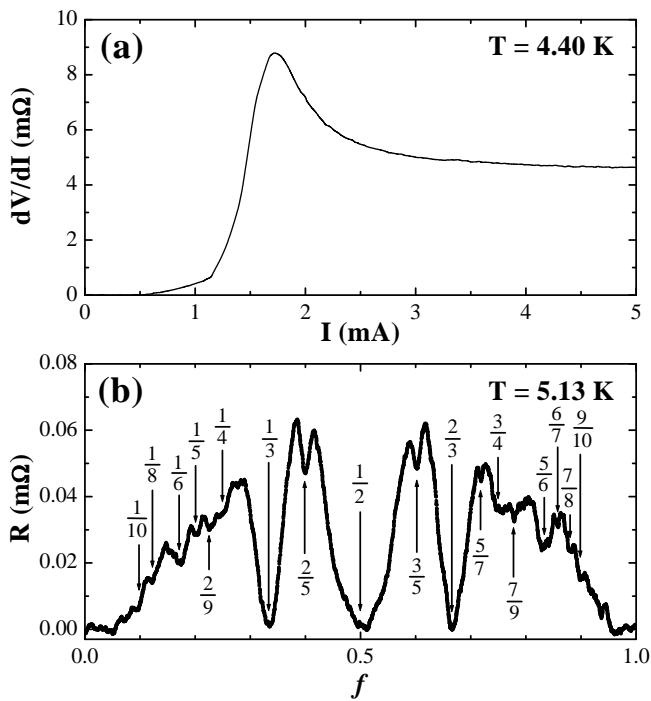}
\vskip 0.1true cm 
\caption {(a) $I$ vs $dV/dI$ at $T = 4.40$ K. (b) Sample resistance with 10-$\mu$A excitation current as a function of frustration at $T=5.13$ K.}
\label{fig1}
\end{figure}

\begin{figure}
\includegraphics[width=1.0\linewidth]{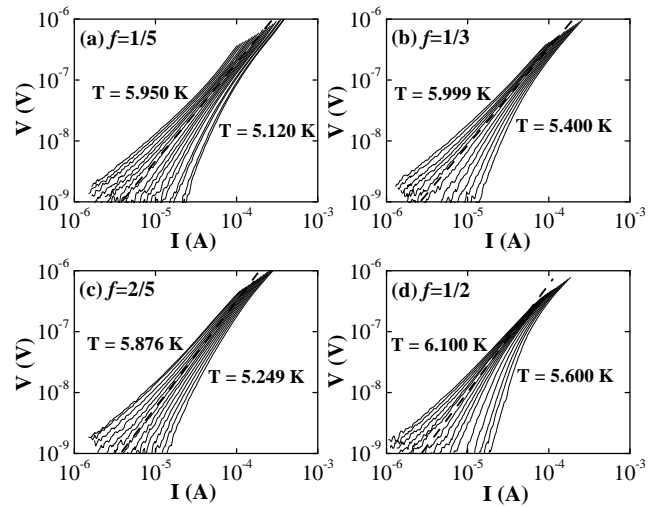}
\vskip 0.1true cm 
\caption{Some of the $IV$ curves for four different frustrations.
(a) $f = {1/5}$ at $T$ = 5.120 to 5.950 K;
(b) $f = 1/3$ at $T$ = 5.400 to 5.999 K;
(c) $f = {2/5}$ at $T$ = 5.249 to 5.876 K;
(d) $f = {1/2}$ at $T$ = 5.600 to 6.100 K.
The dashed lines are drawn to show the power law ($V \sim I^{z+1}$) behavior at the critical temperature.}
\label{fig2}
\end{figure}

The sample studied consists of an array of 200$\times$1000 Nb/Cu/Nb Josephson junctions.
The 0.2-$\mu$m-thick cross-shaped Nb islands patterned on top of a 0.3-$\mu$m-thick Cu film form a square array with a lattice constant of 10 $\mu$m and the junctions are 4 $\mu$m wide and 1.3 $\mu$m long.
The zero-field superconducting transition temperature of the sample $T_{KT}=$ 6.24 K and the width $\sim$ 0.4 K.
The deviation of the single-junction critical current $i_{c}$ or the junction coupling strength $J$ ($=\hbar i_{c} / 2e$) of the sample, estimated from the $I$ vs $dV/dI$ curve in Fig.\ \ref{fig1}(a) and the junction separation fluctuations, was $\lesssim$ 15 \%.
The appearance of higher-order dips in the magnetoresistance curve of the sample with 10-$\mu$A excitation current, shown in Fig.\ \ref{fig1}(b), indicates the good uniformity of the applied transverse magnetic field over the sample.
From the magnetoresistance curve of the sample with 50-$\mu$A excitation current, showing sharp resistance minima at fractional $f$'s, a precise adjustment of the magnetic field or the frustration was possible.
The temperature during the measurements was controlled with fluctuations $\lesssim$ 1 mK.
The standard four-probe technique was adopted for measurements of the $IV$ characteristics.
The voltage $V$ was measured by the use of a transformer-coupled lock-in amplifier with a square-wave current at 23 Hz.
The single-junction critical current $i_c$ at low temperatures can be obtained from the $I$ vs $dV/dI$ curves.
The $i_c$'s at high temperatures were determined by extrapolating the $i_c$ vs $T$ data at low temperatures by the use of de Gennes expression for proximity-coupled junctions in the dirty limit, $i_{c}(T) = i_{c}(0)(1-T/T_{co})^{2}\exp(-\alpha T^{1/2})$, where $T_{co}$ is the BCS transition temperature.

\begin{figure}[t]
\includegraphics[width=1.0\linewidth]{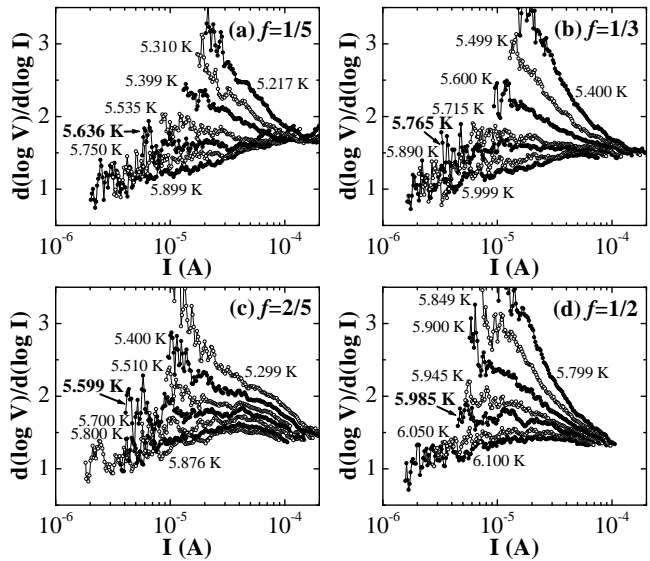}
\vskip 0.1true cm 
\caption{Current dependence of the slope of the $IV$ curves in Fig.\ \ref{fig2}.}
\label{fig3}
\end{figure}

Figure\ \ref{fig2} shows some of the $IV$ traces for $f=1/5$, $1/3$, $2/5$, and $1/2$.
The $IV$ curves at high temperatures are concave upward.
As the temperature is lowered, the curves become straight and then bend progressively more downward \cite{R10}.
The temperature dependence of the $IV$ curves suggests that for all the $f$'s studied, the sample experiences a superconducting transition at the temperature where the straight $IV$ curve appears.
If the superconducting transition is continuous, the $IV$ data are expected to satisfy the scaling relation in 2D, $V/I|T-T_{c}|^{z\nu} = {\cal E}_{\pm}(I/T|T-T_{c}|^{\nu})$, where ${\cal E}_{\pm}$ are the scaling functions above and below $T_c$ \cite{R11}.
This scaling form becomes a simple power-law $IV$ relation, $V \sim I^{z+1}$, at $T=T_c$.
Thus we can find approximate $T_c$ and $z$ directly from the $IV$ curves in Fig.\ \ref{fig2}.
Accurate $T_c$, $z$, and $\nu$ can be obtained from the scaling analyses of the $IV$ data.
Though the scaling analysis of $IV$ data is a powerful tool for identifying a continuous superconducting transition and for determining the critical exponents at the transition, it has been recently shown \cite{R12,R13} that special care should be taken not to extract any false information from the scaling analysis.
A good collapse of $IV$ curves is not sufficient evidence for the transition.
The proposed necessary criterion for $IV$ data supporting a superconducting transition is that log $V$ vs log $I$ isotherms equally distanced from $T_c$ must have opposite concavities at the same applied currents \cite{R12}.
Figure\ \ref{fig3} shows the current dependence of the derivative of the log $V$ vs log $I$ isotherms in Fig.\ \ref{fig2}.
The data at high currents are due to the single-junction effect ($f=1/5$) or the free flux flow ($f=1/3$, 2/5, and 1/2) and thus should be excluded from scaling.
The derivative plots clearly show that the isotherms near $T_c$ have the derivative nearly independent of the current and that the isotherms at temperatures equally distanced from $T_c$ have opposite concavities at the same current level.
For all four $f$'s, our data appear to meet the criterion.

\begin{figure}[b]
\includegraphics[width=1.0\linewidth]{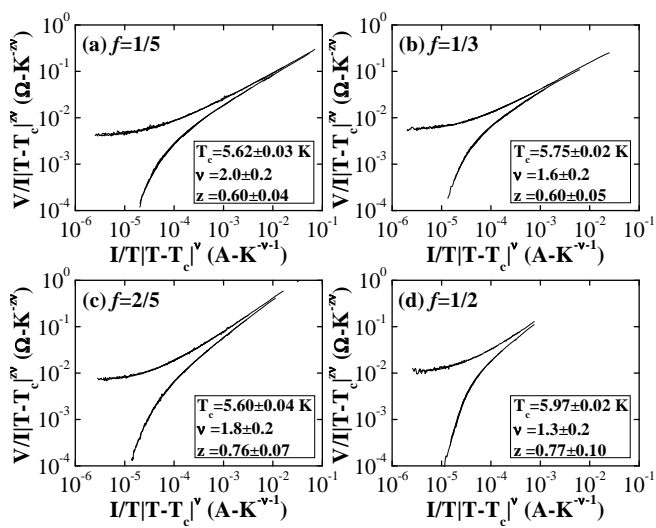}
\vskip 0.1true cm 
\caption{Scaling plots of the $IV$ curves. The values of $T_c$, $\nu$, and $z$ used to scale the data are shown in the insets.}
\label{fig4}
\end{figure}

The $IV$ data scaled on the basis of the scaling form are shown in Fig.\ \ref{fig4}.
Each plot contains $IV$ curves at 16-24 different temperatures.
We find the four sets of data exhibit good scaling behaviors, which suggests the continuous phase transition for all the $f$'s.
As discussed earlier, the continuous transitions at $f$ = 1/5 and 2/5 for which a first-order phase transition is expected \cite{R7,R14} are probably due to the random bond disorder in the sample.
In the scaling analyses, the parameters $T_c$, $\nu$, and $z$ were carefully adjusted until a good collapse of the curves was achieved.
The insets of Fig.\ \ref{fig4} exhibit the values of the parameters used to scale the data. Table\ \ref{table1} summarizes the results of the scaling analyses together with the results of numerical studies of the $XY$ model without disorder \cite{R5,R7,R8} in parentheses for comparison.
Both $T_c$ and $\nu$ from the $IV$ measurements are higher than the numerical predictions in the absence of disorder.
We note that for all the $f$'s including $2/5$ and $1/3$, the critical exponent $\nu$ considerably exceeds the 2D Ising value ($\nu=1$).
It is interesting that the $\nu$ ($= 1.8$) for $f=2/5$ is close to the SWN value (1.8 - 1.9) at the near-irrational frustration $f=0.618\simeq (\sqrt{5}-1)/2$ \cite{R6,R9}, the complementary fraction ($1-f$) of which is 0.382 in the vicinity of 2/5.
The results of critical scaling analyses of the $IV$ data indicate that quenched bond disorder may drive the phase transitions of frustrated JJA's to be continuous but not into the Ising universality class, contrary to the observed in simulations at $f=2/5$ with disorder and at $f=1/3$.

\begin{table}[b]
\caption{$T_c$, $\nu$, and $z$ from the scaling analyses of the $IV$ data.
The numbers in parentheses are those from numerical studies of the $XY$ model without disorder \cite{R5,R7,R8}.
The values of $T_c$ are in unit of $J/k_B$.}
\label{table1}
\begin{ruledtabular}
\begin{tabular}{llll}
$f$     & $T_c$ & $\nu$ & $z$ \\
\hline
$1/5$ & 0.23 ~(0.18) & 2.0 & 0.60 \\
$1/3$ & 0.30 ~(0.2185) & 1.6 ~(0.993) & 0.60 \\
$2/5$ & 0.22 ~(0.2127) & 1.8 & 0.76 \\
$1/2$ & 0.46 ~(0.451-0.455) & 1.3 ~(0.80-1) & 0.77 \\
\end{tabular}
\end{ruledtabular}
\end{table}

It was argued \cite{R5} that the Ising-like transitions of the $XY$ model at $f=1/3$ and at $f=2/5$ with disorder are because of domain wall excitations with configurations of the Ising model.
The arguments suggest that the excitation configurations of the JJA's in the $IV$ measurements must be different from those of the $XY$ systems in the simulations.
The larger $\nu$'s for JJA's imply that the JJA has a much longer correlation length near $T_c$ than the corresponding $XY$ system.
For the present, we have no plausible explanation for the disagreement between experiments and simulations.
We speculate only about the disagreement.
Though JJA's are known to provide a near-ideal implementation of the $XY$ model, JJA's in the $IV$ measurements could differ in the following aspects from the $XY$ model in the simulations.
In the experiments, the critical behavior was studied on JJA's with a current injected.
If an injected current alters fundamental ingredients of the system, the nonequilibrium measurements might bring out the critical exponents different from those in equilibrium.
On the other hand, in the simulations, the periodic boundary condition was imposed to compensate the effects of a finite size of the model system investigated.
It has been found in MC simulations of the 2D $XY$ model near incommensurability \cite{R15} that when the system has a natural period different from the imposed period, the periodic boundary condition induces numerous extra low-energy walls, which may have a significant effect on transitions in smaller systems.
If it is also the case for $f=1/5$, $1/3$, $2/5$, and $1/2$, the critical exponents from the simulations at the $f$'s could be different from those of real arrays with larger sizes and free boundaries.

Another remarkable feature of our data is that the dynamic critical exponent $z$ of the JJA's is only 0.60 - 0.77, much smaller than that ($z=2.0 \pm 0.5$) observed \cite{R6} in SWN's which are known in the same static universality class as JJA's.
This implies that in the presence of a magnetic field, JJA's relax near $T_c$ much more rapidly than SWN's.
The JJA values for $z$ stand closer to the value from Langevin-dynamics simulations ($z \approx 1$ for $f=1/2$) \cite{R16} than to the SWN value.
The Langevin-dynamics simulations of the resistively shunted-junction model and of the resistively capacitively shunted-junction model show that the $IV$ characteristics of a JJA for $f=1/2$ are dominated by the motion of domain walls rather than the motion of vortices.
$z\simeq2$ is seen at the transition of the frustrated $XY$ model governed by the purely dissipative dynamics (without conservation of order parameter) \cite{R17} as well as at a Kosterlitz-Thouless-Berezinskii transition.
We speculate that such low $z$'s for the JJA's are possibly related to the brisk motion of domain walls.
It is, however, not clear whether $z\simeq2$ for SWN's is from stronger dissipations or from some other origins.

In summary, the scaling behavior of the $IV$ characteristics of a square JJA with quenched bond disorder was investigated for $f = 1/5$, $1/3$, $2/5$, and $1/2$.
For all the frustrations, a good scaling behavior of the $IV$ characteristics was found.
The critical exponent $\nu$ exceeding the 2D Ising value suggests that bond disorder may drive the transitions of JJA's continuous but not to the 2D Ising universality class, contrary to what was observed in MC simulations of the $XY$ model.
The dynamic critical exponent $z$ as low as 0.60 - 0.77 indicates that the frustrated JJA's relax rapidly in the vicinity of $T_c$.

This work was supported by the BK 21 program of the Ministry of Education.


\begin{references}
\bibitem {R10}
The change in the $IV$ characteristics caused by the finite sample size, which has been found in the absence of a magnetic field serious enough to alter the scaling parameters considerably, is not significant in the presence of a magnetic field.
For our sample with $I=10^{-5}$ A at $T=6.0$ K, $V(f=1/3) \approx 2\times10^{-8}$ V, whereas $V(f=0) < 10^{-10}$ V, a part of which is due to the finite-size-induced vortices.
It is probably because, in the presence of a magnetic field, the field-induced vortices greatly outnumber the finite-size-induced one.
\end{references}
\end{document}